\documentstyle[12pt,aasms4,flushrt]{article}

\newcommand{\nc}{\newcommand}
\nc{\bec}{\begin{center}}
\nc{\enc}{\end{center}}
\nc{\beq}{\begin{equation}}
\nc{\enq}{\end{equation}}
\nc{\bei}{\begin{itemize}}
\nc{\eni}{\end{itemize}}
\nc{\bee}{\begin{enumerate}}
\nc{\ene}{\end{enumerate}}
\nc{\beqar}{\begin{eqnarray}}
\nc{\enqar}{\end{eqnarray}}
\nc{\namely}{{\it viz.}}
\def\lsun{$L_\odot${}}
\def\micron{\hbox{$\mu$m}}
\def\msun{$M_\odot$}
\def\dep{$\tau_{\rm 100}$}

\lefthead{MOOKERJEA et al. }
\righthead{FIR observations of IRAS 00338+6312 \& RAFGL 5111}

\begin{document}


\title{Far Infrared Observations of the Galactic Star Forming Regions associated with IRAS 00338+6312 \& RAFGL 5111}

\author{ B. Mookerjea \altaffilmark{1}, S.K. Ghosh\altaffilmark{1}, A.D. Karnik\altaffilmark{1}, T.N. Rengarajan\altaffilmark{1},\\ S.N. Tandon \altaffilmark{2} \& R.P. Verma\altaffilmark{1} }

\altaffiltext{1}{Tata Institute of Fundamental Research,
              Homi Bhabha Road,  
              Mumbai  400 005, India}
\altaffiltext{2}{Inter-University Centre for Astronomy \& Astrophysics,
              Ganeshkhind, Pune 411 007, India}


\begin{abstract}

  Two Galactic star forming regions, one in a very 
early phase of evolution and another evolved one,
associated with the
IRAS sources 00338+6312 and 03595+5110 (RAFGL 5111) respectively
have been studied in detail. 
These sources have been mapped simultaneously in two
far infrared bands ($\lambda_{\rm eff}$ = 143 \& 185 \micron),
with $\sim$ 1\farcm 5 angular resolution, 
using the TIFR 100 cm balloon borne telescope. 
The HIRES processed IRAS maps
at 12, 25, 60 \& 100 \micron, have been used for comparison.
Whereas IRAS 00338+6312 is resolved only in the TIFR bands, RAFGL 5111
is very well resolved in both the TIFR bands, as well as
in at least 3 IRAS bands. The neighbouring fainter source 
IRAS 04004+5114 has also been resolved in the TIFR bands.
Taking advantage of the identical
beams in the two TIFR bands at 143 \& 185 \micron, dust
colour temperature, $T(143/185)$, and optical depth, $\tau_{\rm 150}$,
maps have been generated for RAFGL 5111. 
These maps show interesting structural details. 

Radiative transfer modelling in spherical geometry has been carried out 
for individual sources to extract information about:
the cloud size, type of the embedded source, 
radial density distribution, optical depth, gas to dust ratio,
and the dust grain composition.  The best fit models are in 
good agreement with the observed spectral energy distribution (SED),
radio continuum data etc.
Another scheme of radiative transfer through the interstellar 
dust-gas cloud including the heavier elements has been used 
to predict ionic nebular line emission, 
which are in reasonable agreement with the measurements for RAFGL 5111.
 An important conclusion from the present study is that, for all the three
sources (IRAS 00338+6312; 03595+5110; and 04004+5114, a faint 
source in the neighbourhood of RAFGL 5111), the best fit to the
observed SED is obtained for a uniform density ($n(r) \sim r^0$) cloud.

\end{abstract}

{\em Subject Headings}: { IRAS 00338+6312 -- RAFGL 5111 -- IRAS 03595+5110
-- IRAS 04004+5114 -- Far Infrared Mapping -- Radiative Transfer
-- H II region}



\section{Introduction}

In order to get complete information and understanding 
of the energetics,
physical sizes, spatial distribution of interstellar dust,
and the role of the cooler ($<$ 20 K) dust in Galactic 
star forming regions, high angular resolution mapping 
in trans-IRAS wavebands ($\lambda_{eff} >$ 100 \micron)
is essential. With the above in view, 
a programme of high angular resolution mapping 
of Galactic star forming regions in far infrared (FIR) bands
using the TIFR 100 cm balloon borne telescope, was undertaken. 
At present a two band FIR photometer, with effective wavelengths of
143 \micron\ and 185 \micron\ 
is in regular use. This photometer uses bolometer arrays 
thereby resulting in much higher observational efficiency 
(Verma, Rengarajan \& Ghosh 1993). 
Several individual Galactic star forming regions have been
observed in a series of balloon flights during 1993-98 
(Ghosh et al. 1996, Karnik et al. 1999, Mookerjea et al. 1999).
The present paper deals with the two
star forming regions associated with IRAS 00338+6312 and 
03595+5110 (RAFGL 5111). 

 IRAS 00338+6312 is an unidentified IRAS Point Source Catalog
(hereafter IRAS PSC) source with its spectrum rising steeply
from 25 to 60 to 100 \micron, typical of YSO's.
It is located near the core of the dark cloud L1287. 
A near-IR survey of Herbig-Haro objects and
candidates (with red nebulosity) led Persi et al. (1988) to
suggest that RNO 1 may be identified with IRAS 00338+6312.
Snell, Dickman \& Huang (1990) discovered an energetic 
molecular outflow activity in IRAS 00338+6312 from their 
${}^{12}$CO measurements. Further millimeter wave line studies
in CO, HCN and HCO$^{+}$ by Yang et al. (1991) confirmed
the outflow activity. They identified the driving source for the
outflow to be IRAS 00338+6312.

 With the discovery of a new FU Orionis type source, RNO 1B
(which brightened by over 3 magnitudes since 1978), right inside
the IRAS positional error ellipse, Staude \& Neckel (1991)
suggested that the energetic outflow is powered by FU Ori
activity. This was supported by the discovery of a second
FU Orionis source RNO 1C (Kenyon et al. 1993); \& CS line
and mm-wave continuum study of the FU Ori binary system
RNO 1B/1C, by McMuldroch, Blake \& Sargent (1995).

  The resolution to the debate between whether the outflow
activity is powered by a deeply embedded YSO or FU Ori
activity, came from the following clinching evidences :
(i) high spatial resolution (0.1\arcsec ) near-IR (3.8 \micron) 
polarimetric mapping by Weintraub et al. (1996) found 
the centroid of the polarization vectors to 
lie within 1\arcsec\ of the IRAS position;
(ii) the discovery of radio continuum counterpart to IRAS
00338+6312 at 3.6 cm by Anglada et al. (1994), named ``VLA 3'',
again within 1\arcsec\ of IRAS coordinates;
(iii) additionally, Henning et al. (1992) detected a H$_{2}$O
maser associated with IRAS 00338+6312.
The above clearly indicated that FU Oris (RNO 1B/1C) 
are not physically associated with the outflow activity/
IRAS 00338+6312, but a deeply embedded YSO at the core of L1287
is energizing the source.

   The second object of the present study, IRAS 03595+5110 is a
strong far infrared source with its spectrum rising from  60
to 100 \micron. The IRAS PSC itself identified it with sources 
in 7 catalogues including AFGL, catalogues of: CO sources, 
H II regions and bright Galactic nebulae. 
The extended nature of IRAS 03595+5110 at mid and far infrared
wavelengths led it to be included in the IRAS Small Scale
Structure Catalog (X0359+511). This
source is well inside the nebulosity of Sharpless source S 206,
an evolved H II region, which has been 
studied extensively in optical and radio wavelengths. The
evolved nature of this source is also supported by the fact 
that although strong CO molecular line emission has been
found (Snell, Dickman \& Huang, 1990), no outflow activity
was detected. The most outstanding feature about S 206 is
its geometry: it is an example of a ``blister''  type of
H II region, i.e. a dense molecular cloud being illuminated
/heated/ionized by a star 
positioned outside the cloud itself. Our line of sight
to S 206 is roughly normal to the shortest line joining
the exciting star and the outer surface of the molecular cloud.
The region around IRAS 03595+5110 has been an important
target for the following studies : mapping in near infrared (Pismis
\& Mampaso 1991) and radio continuum (Albert et al. 1986,
Balser et al. 1995); and H$_{2}$O maser search (Henning et al.
1992).

Although the two sources chosen for the present study  
are positionally located close to nebulosities, one of them,
IRAS 03595+5110,
is genuinely associated with the nebula (an example of a very
evolved star forming region), and the other 
one, IRAS 00338+6312, accidentally
happens to be near the line of sight to the FU Oris, but 
actually represents an extremely young star forming region.

\section{Observations}

The two Galactic star forming regions, \namely ,  IRAS 00338+6312 and
RAFGL 5111, have been observed using the two band far infrared (FIR)
photometer system at the Cassegrain focus of the  TIFR 100 cm (F/8) balloon 
borne telescope.  The FIR telescope was flown
from the TIFR Balloon Facility, Hyderabad, in central India (Latitude
= 17\fdg 47 N , Longitude = 78\fdg 57 E ), on November 12, 1995.  
Details of the telescope and the
observational procedure are given by Ghosh et al. (1988). Only a brief
description is presented here. 
The two FIR bands use a pair of 2 $\times$ 3 composite Silicon 
bolometer arrays 
cooled to 0.3 K by liquid $^{3}$He, which view
identical part of the sky simultaneously.  The field of view
of each bolometer is 1\farcm 6. 
The sky is chopped along the cross-elevation axis at 10 Hz 
with a throw of 4\farcm 2. 
The two FIR wavebands are defined by a cooled beam splitter 
(restrahlen filter) and other filters. 
The normalised transmission curves for the two bands of the 
FIR photometer are shown in Fig. 1, which have been measured
in the laboratory using a Michelson spectrometer 
with a Golay cell as the reference detector. 
The effective wavelengths for the
two bands, for a $\lambda ^{-1}$ emissivity law
and $\sim 36$ K temperature are, 143 \micron\ and 185 \micron. 
Saturn was observed for absolute
flux calibration as well as determination of the instrumental Point Spread
Function (PSF) including the effect of sky chopping.

The simultaneous mapping in the two FIR bands was carried out by
raster scanning the telescope along the cross-elevation axis 
across the target source area under study.
Whereas IRAS 00338+6312 was covered by three rasters, RAFGL 5111
was covered twice. The chopped FIR signals were gridded in
a two dimensional sky matrix (elevation $\times$ cross elevation), with
a cell size of 0\farcm 3$\times$ 0\farcm 3 . 
The observed signal matrix was 
deconvolved using an indigeneously developed procedure using the
Maximum Entropy Method (MEM) similar to Gull
and Daniell, 1978 (see Ghosh et al., 1988, for details). 
The FWHM sizes  of the
deconvolved maps of the point-like source (Saturn)
are 1\farcm 6$\times$1\farcm 9  and 1\farcm 6$\times$1\farcm 8  
in the 143 and 185 \micron\ bands respectively.
For improving the offline absolute aspect determination of the telescope,
a pre-selected star ($m_{B} < $ 8) in a ``clean'' field located within 
30\arcmin\ of the target FIR source, was mapped using an optical
photometer. The optical maps were also generated by using the same
2-D MEM deconvolution procedure as used for the FIR signals.
The resulting absolute positions of the maps are good to $\sim$ 1\arcmin\ .
The positions of the stars detected by the optical photometer 
(during the raster scans across the FIR target source), are used to correct 
for any systematic shifts.  This procedure leads to an 
absolute positional accuracy in the FIR maps, to about 0\farcm 5 .

\section{The HIRES processed IRAS maps}

  The IRAS survey data at all four bands (12, 25, 60 \& 100 \micron)
for the relevant regions of the sky corresponding to IRAS 00338+6312
and RAFGL 5111 were HIRES processed (Aumann, Fowler \& Melnyk, 1990) at 
Infrared Processing and Analysis Center (Caltech).\footnote[1]{IPAC is
funded by NASA as part of the part of the IRAS extended mission
program under contract to JPL.}
These maps for both the programme sources have been used in
the present study to quantify flux densities and angular sizes at the
four IRAS bands. They have also been used for generating temperature
and optical depth maps when the source is resolved.

\section{Results}

\subsection{IRAS 00338+6312}

An area of 22\arcmin $\times$7\arcmin\ centred on 
IRAS 00338+6312 was mapped in the two FIR bands at 143 and 185 \micron.  
The deconvolved TIFR maps at these two bands 
are presented in Fig. 2. 
The dotted lines in these figures, mark the boundary of the region 
covered by the telescope bore-sight. 
However, the actual observations extend beyond this
boundary due to : (i) the size of the detector arrays;
(ii) the sky chopping; \& (iii) the effective PSF size.
As a result, the MEM deconvolved intensity maps
extend $\sim \pm$ 6\arcmin\ beyond the marked boundary
along each axis.The insets show the 
deconvolved intensity contours for the point-like source (Saturn).
Since the 185 \micron\ map has higher dynamic range, 
the isophot contours are shown upto 2.5\% of the peak intensity 
(809 Jy/sq. arcmin), while 
at 143 \micron\ the isophot contours have been displayed
only upto 10\% level of the peak intensity (368 Jy/sq. arcmin),

The HIRES processed maps at all the four
IRAS bands, for the source IRAS 00338+6312 are shown 
in Fig. 3.
The peak intensities in these 12, 25, 60 \& 100 \micron\
maps correspond to 8.2, 82.8, 312 \& 284 Jy/sq. arcmin
respectively. Although the dynamic range in all these
maps is much larger, for the purpose of comparison
with TIFR maps, the isophots have been shown upto 1\%
level of respective peak intensities.
 The angular resolutions achieved in these maps 
(In-Scan $\times$ Cross-Scan) 
are 0\farcm 45 $\times$0\farcm 93 , 
0\farcm 47$\times$0\farcm 72  ,
0\farcm 78$\times$1\farcm 37  and 1\farcm 65$\times$
2\farcm 05 at 12, 25, 60 and 100 \micron\ respectively.

  The positions of the peaks corresponding to IRAS 00338+6312 in 
143 \& 185 \micron\  maps, are shifted by 12\arcsec\ 
\& 33\arcsec\ respectively
with respect to the IRAS PSC coordinates. Considering the 
absolute positional accuracy of our maps ($\approx$ 30\arcsec\ )
and errors in the IRAS coordinates, they are consistent.

The results of numerical aperture photometry on the peaks in all 
six maps, are presented in Table 1.
  The flux densities extracted from the HIRES maps (within 5\arcmin\
dia circle) are 1.4, 1.1, 1.1 \& 1.2 times the IRAS PSC values
at 12, 25, 60 \& 100 \micron\ bands respectively. 
Interestingly, the uncertainties in the flux densities from
the HIRES maps have been determined to be 24\%, 
7\%, 1\% \& 3\% at the four bands
respectively, which mainly originate from the definitions of
the local background levels (leading to overestimation). 
In addition, considering that the HIRES flux densities
represent slightly larger solid angle than the PSC, it may
be concluded that the PSC and HIRES data are consistent
and IRAS 00338+6312 is unresolved by IRAS at all 4 bands.

The IRAS 00338+6312 has been resolved in both 143 \& 185 \micron\ bands,
but with a morphological difference. Whereas in the 143 \micron\
map, the FWHM size clearly shows the extended structure,
(1\farcm 9  $\pm$0\farcm 1$\times$2\farcm 0, compared to 
1\farcm 6$\times$1\farcm 9  for Saturn);
at 185 \micron\  the extension is clearly visible only at
lower contour levels of 10 \% (3\farcm 0$\times$3\farcm 1, compared to
2\farcm 5$\times$2\farcm 8  for Saturn) and 
5 \% (3\farcm 5$\times$3\farcm 6, compared
to the 2\farcm 8$\times$3\farcm 2  for Saturn). 
These angular dimensions contain information regarding the dust 
temperature distribution around IRAS 00338+6312.
Interestingly, the extension is in a direction perpendicular to the molecular
outflow direction or the extension seen in the radio continuum 
map.

Although IRAS 00338+6312 has been resolved in both 143 \&
185 \micron\ bands, the structural information is rather limited
for generating meaningful spatial distribution of colour 
temperature/optical depth.
Instead, an attempt has been made to construct a radiative
transfer model of this source, which is consistent with 
all available observational data.
The flux densities from the present study along 
with other data available from
the literature, have been compiled to generate
the Spectral Energy Distribution (SED) for IRAS 00338+6312,
to constrain the model (Section 5.2).

\subsection{RAFGL 5111 (IRAS 03595+5110)}

For RAFGL 5111 (IRAS 03595+5110), an area of 
24\arcmin $\times$11\arcmin\  around this source was mapped in the two
FIR bands. The deconvolved TIFR maps for RAFGL 5111 at 143 \& 185
\micron\ are presented in Figs. 4 (a) \& (b) respectively.
The isophots in both the maps 
have been displayed upto the 10\% level of 
the respective peak intensities. The peak intensities are
142 Jy/sq. arcmin at 143 \micron\ band and 
99.2 Jy/sq. arcmin at 185 \micron\  band.

IRAS 03595+5110 (RAFGL 5111) has been resolved 
very well in both the TIFR
maps, and the neighbouring source IRAS 04004+5114 has also
been detected very clearly in both the bands, with 
some structural details too.

  The positions of the global peaks 
corresponding to IRAS 03595+5110 
in our 143 \& 185 \micron\ maps, are shifted with
respect to the IRAS PSC coordinates by 0\farcm 3 and 0\farcm 5 
respectively.
Considering our absolute positional accuracy, they are
consistent. However, the position of the fainter
nearby source corresponding to IRAS 04004+5114, is quite different
in 143 \& 185 \micron\ maps from the IRAS PSC coordinates. 

The HIRES processed IRAS maps (upto 1\% level of the peak)
for the corresponding region around IRAS 03595+5110 are shown 
in Figs. 4(c) -- (f). The peak intensities in the HIRES maps are 9.66, 
93.4, 237 and 130
Jy /sq. arcmin at 12, 25, 60 \& 100 \micron\ bands respectively.
The angular resolutions achieved in these maps are
0\farcm 45$\times$1\farcm 05, 
0\farcm 47$\times$0\farcm 83,
0\farcm 85$\times$1\farcm 28 and 
1\farcm 77$\times$2\farcm 30 
at 12, 25, 60 and 100 \micron\ respectively.
At least in the 12, 25 and 60 \micron\ maps, IRAS 03595+5110
is very well resolved and at 100 \micron\ there is some indication
of extension. The nearby source IRAS 04004+5114 is also seen
clearly in all the four bands (though at 100 \micron, appropriate
levels of the contours need to be chosen, to be able to 
see the peak corresponding to IRAS 04004+5114, in the presence of
the much stronger emission from IRAS 03595+5110).

  The ratios of the integrated flux densities in a 5\arcmin\ diameter
circular aperture around IRAS 03595+5110 in the HIRES maps,
to the corresponding IRAS PSC values are 11.3, 5.4, $>$ 3.2
and 1.8 at 12, 25, 60 \& 100 \micron\ bands respectively. This
quantifies the extended nature of this source in these bands.
All the flux density and position details are given in Table 1.

Although the IRAS (HIRES) maps have much higher dynamic range,
the angular resolution of TIFR maps are superior to the IRAS
maps (at least at 60 \& 100 \micron), due to the smaller and circular
beams employed.
Since the TIFR beams at both the FIR bands are identical and all
the observations are simultaneous, this dataset is very sensitive
to detect gradients in colour temperature and/or dust optical depth
in this source. Hence,
we have generated maps for the colour temperature as well as for the 
optical depth at 150 \micron, assuming a dust emissivity law of 
${\epsilon}_{\lambda} \sim \lambda^{-1}$. 
These colour temperature and optical depth maps, have been 
generated, by using an interpolation table relating the ratio of 
signals detected for the two bands
to the dust temperature for the assumed emissivity law.

  Since the 143 and 185 \micron\ maps in general have different
dynamic ranges, the colour temperature/optical depth maps
have been generated only for that region around IRAS 03595+5110,
where the intensities in each map are above corresponding
prescribed thresholds. These thresholds reflect the achieved
dynamic ranges in the respective maps (which are 5 \& 10 at
143 \& 185 \micron\ bands). In addition, to be very
conservative regarding claiming structural information in the
temperature/optical depth maps, further averaging (over 
3$\times$3 pixels)
of both the intensity maps at 143 \& 185 \micron, have been carried out
before computing the temperature/optical depth.
A region about 5\arcmin\ $\times$  4\arcmin\
in size around IRAS 03595+5110 (RAFGL 5111) qualifies the above criteria.
The resulting colour temperature map, $T(143/185)$, is shown in 
Fig. 5 (a). 
The dust colour temperature distribution shows interesting structure 
with one major peak and an indication of a hotter region to the east
of our map ($\sim$ 2\arcmin\ east of  the intensity peak). The temperature
ranges between 20 and 43 K.
The hottest dust is systematically offset 
to the west (by $\approx$ 1\farcm 1  ),
from the peak position of the intensity maps.
Such offsets have been predicted by ``blister'' type models
of H II regions, where the exciting star lies near the 
edge/outside the molecular cloud (Icke, Gatley, \& Israel, 1980).
Hence, our dust temperature map further supports the ``blister''
type geometry for RAFGL 5111, which has been suggested 
in the literature from radio continuum maps (Albert et al. 1986).

The dust optical depth map of the same region at 
150  \micron, $\tau_{\rm 150}$, (for assumed emissivity law of
${\epsilon}_{\lambda} \sim \lambda^{-1}$) is shown in Fig. 5 (b). 
The $\tau_{\rm 150}$ map shows a peak (2.6$\times$10$^{-3}$) towards
the east of the intensity peak and a negative gradient terminating
in a minimum (5.2$\times$10$^{-4}$) towards the west. The minimum
value of $\tau_{\rm 150}$ positionally corresponds to the maximum
of $T(143/185)$.

  Although the HIRES maps at 60 and 100 \micron\ have relatively
poorer angular resolution compared to the TIFR far infrared maps,
they have been used to generate the colour temperature, $T(60/100)$,
and optical depth ($\tau_{100}$) maps (see  Figs. 5(c) \& (d)) for 
the same region as in Figs. 5(a) \& (b) . The intensity maps at 
60 \& 100 \micron\
were averaged over 1\arcmin $\times$ 1\arcmin\ before 
computing $T(60/100)$ and $\tau_{100}$ in a manner similar to 
that described by Ghosh et al. (1993) for  
an emissivity law of ${\epsilon}_{\lambda} \sim \lambda^{-1}$.
The $T(60/100)$ map shows one major peak (67 K), slightly to the 
west and two minor peaks (46 K) to the east ($\sim$ 2\arcmin) of the 
intensity peak.
It is interesting to compare the $T(60/100)$ map with the $T(143/185)$
map. Both are very similar morphologically. In both the maps,
the hottest dust is positioned to the west of the intensity
peak corresponding to IRAS 03595+5110. This westward offset 
is larger for the $T(143/185)$ map. This is again consistent with
the predictions 
for an H II region with ``blister'' type geometry (Icke, Gatley \&
Israel, 1980).
In addition, a local enhancement in dust temperature is seen in both 
$T(60/100)$ and $T(143/185)$ maps at the same position ($\sim$ 2\arcmin\
east of IRAS 03595+5110).
The \dep\ map shows two peaks, one to the east (10$^{-3}$) and 
other to the west 
(1.2$\times$10$^{-3}$) of IRAS 03595+5110. The region between 
these peaks, around the IRAS source position, show a uniform value
of \dep. The only peak in $\tau_{\rm 150}$ map coincides with the
eastern peak of \dep. The ratio ``$r$'' ($\tau_{\rm 150}$/\dep) 
for the peak position (after correcting for $\lambda^{-1}$
emissivity) is $\sim$ 3.9. The other (western) peak in \dep\ map is
coincident with the minimum in $\tau_{\rm 150}$, with $r$ being 1.5.
In the intervening region, $r$ takes up intermediate values. The above
may not be surprising, since the IRAS and TIFR maps are
probing dust at different temperatures. The spatial variation in
$r$, along the east-west line through IRAS 03595+5110, could
indicate temperature and/or density gradient as expected in
``blister'' type geometry in this source.

\subsubsection{IRAS 04004+5114}

 In order to appreciate the structural details of the
fainter source (neighbouring RAFGL 5111),
i.e. IRAS 04004+5114, the relevant regions of the
TIFR (143 \& 185 $\mu$m) as well as IRAS HIRES (60 \& 100  \micron)
maps have been displayed in Fig. 6. 
For generating these 143 \& 185 $\mu$m maps, the observed
signals corresponding to the nearby stronger
source (IRAS 03595+5110) have been ``masked'' ({\it i.e.} as if no
measurements are available at those locations)
before carrying out the MEM deconvolution.
The contour levels in all the four maps
have been chosen appropriately to highlight the morphology
of IRAS 04004+5114.
The peak intensities in 143, 185, 60 \& 100 \micron\ bands are
34.3, 31.7, 22.7 and 23.8 Jy /sq. arcmin respectively.
Whereas the TIFR maps are shown to 40\% of the respective peaks
due to limited dynamic range,
the HIRES maps contain structural information upto 1\%
of the corresponding peaks.
Most of the approximately east-west (P.A. $\approx$ 75$^{\circ}$)
extension seen for IRAS 04004+5114 in the
HIRES 60 \& 100 \micron\ maps, could have been explained by the
scan track of the IRAS focal plane (IRAS detectors are rectangular
with longer side in the cross-scan direction). But, even
in the TIFR maps at 143 \& 185 \micron\ (with circular beams), 
this source is extended along a similar position angle.

Flux densities for IRAS 04004+5114 (in all 6 bands)
correspond to an aperture of size 3\arcmin, 
since it is close to the boundary of the TIFR maps (see Table 1).
The ratios of the flux densities obtained from the HIRES
maps to the IRAS PSC values,  for IRAS 04004+5114,
are 2.3, 1.7, 1.5 and 1.4 at 12, 25, 60 \& 100  \micron.
These ratios also indicate that the source is extended.

\section{Discussion}

\subsection{Radiative Transfer Modelling}

Before going into the individual details of the observations of 
IRAS 00338+6312 and the RAFGL 5111 regions, 
in this section we describe the modelling procedure
that has been adopted to interpret the results obtained for each of these
sources. 

\subsubsection{Procedure ``A'': Modelling continuum emission from
dust and gas}

The star forming region has been modelled as a spherically 
symmetric cloud powered by a centrally embedded source,
which could either be a single or a cluster of ZAMS star/(s). 
A schematic diagram of the geometry of the model 
is presented in Fig. 7.
The cloud is assumed to be immersed
in an isotropic radiation field (typical Interstellar Radiation
Field, ISRF) and the interstellar gas consists 
of only hydrogen. The gas and the dust 
follow the same radial density distribution law, but
with the following difference $\--$
whereas the gas exists throughout the cloud (i.e. right from the
stellar surface upto the outer boundary of the cloud, R$_{\rm max}$)
there is a natural lower limit to the inner 
boundary, R$_{\rm min}$, for
the dust distribution (i.e. a cavity in the dust cloud).
This is because the dust grains
are destroyed when exposed to excessive radiative heating.
The gas to dust ratio, where they co-exist ( R$_{\rm min} < $r $<$
R$_{\rm max}$), is assumed to be constant. The position of the
ionization front, R$_{\small \rm HII}$, depends on
the spectral shape and luminosity of the exciting source, 
as well as the density of the gas. The case,
R$_{\small \rm HII}$ $<$ R$_{\rm min}$ is also possible, if either the star 
is not hot enough and/or
the density of gas around the star is quite high. 

Modelling a specific star forming region involves 
matching its predictions to: (i) the observed emergent spectral 
energy distribution (SED) due to thermal emission from dust;
(ii) the radial profiles at various wavelengths; and (iii) the radio
continuum emission from the gas component.
The models are further constrained by the total
luminosity of the embedded energy source/s 
(as determined by integrating the  observed SED).
For all the models the dust component is assumed to be composed of
three most commonly accepted types : Graphite, Astronomical Silicate 
and Silicon Carbide (SiC). 
All physical properties 
of the grains, \namely , absorption and scattering 
efficiencies, have been taken from Draine \& Lee (1984) and 
Laor \& Draine (1993). 
The size distribution of the dust grains is assumed in 
accordance with Mathis, Rumpl \& Nordsiek (1977), to be 
a power law, \namely ,
$n(a)da \sim a^{-m}da$, $a_{min} \leq a \leq a_{max}$
with $m$ = 3.5. The lower and upper limits of the
grain size distribution $a_{min}$ (0.01 \micron) and $a_{max}$
(0.25 \micron) have been taken from Mathis, Mezger \& Panagia (1983). 
The ISRF has been taken from Mathis, Mezger \& Panagia (1983),
and has been held fixed for all model runs.

The following parameters are explored in order
to get an acceptable fit to all the data :
(i) the nature of the embedded source, which could either be a 
 single ZAMS star or a cluster of ZAMS stars consistent with the
 Salpeter Initial Mass Function;
(ii) radial density distribution law (only three power laws
have been explored, \namely , $n(r) \approx r^{0}, r^{-1}$ or
$r^{-2}$);
(iii) the relative abundances of the three constituent grain types;
(iv) total radial optical depth due to the dust (inclusive of all
constituents) at a selected wavelength ($\tau_{100}$ at 100 \micron); 
(v) the gas to dust ratio by mass (the predicted radio 
continuum emission is sensitive to this);
(vi) geometric details like R$_{\rm max}$ \& R$_{\rm min}$ (R$_{\rm min}$
will not violate radiative destruction of grains).

The interstellar cloud is divided into $\approx$ 100 radial grids.
Near both the boundaries, these grids are
logarithmically spaced (in rest of the cloud, a linear grid has
been used). The frequency grid consists of 89 points
covering the wavelength range 944 \AA ~ to 5000 \micron.

  The radiative transport through the dust-gas cloud
has been carried out by using a programme based on the code CSDUST3 
(Egan, Leung \& Spagna 1988). We have improvised this code 
by generalizing the boundary conditions leading to 
much better flexibility for modelling typical
astrophysical sources.
Although originally this code considered only the dust component,
we have modified it to consider and treat the interstellar gas
(only hydrogen) along with dust in a self consistent manner.
With this modification, it has been possible to
predict the radio continuum emission also, using a very simple
approach. This scheme considers photo-ionization and recombination, along
with absorption due to the grains. 
Whereas self-absorption of the radio emission within the
cloud has been considered, the gas-dust coupling has been neglected.
Further details of the procedure are given by Mookerjea \& Ghosh (1998).

For preserving the energetics precisely and self-consistently, 
the total energy available for heating of the dust component,
includes the following three components (all components being binned into
the respectively relevant spectral grid elements) :
(i) the star cluster/ZAMS stellar luminosity in photons 
below the Lyman limit ($\lambda > 912 $\AA);
(ii) a part of the Lyman continuum luminosity of the embedded star, 
($\lambda <912 $\AA), directly absorbed by the dust; and
(iii) a fraction of the same reprocessed by the gas.
The last contribution, \namely , the reprocessed Lyman continuum 
photons, has been quantified by the prescription of
Aller \& Liller (1969) that each Lyman 
continuum photon emitted by
the star ultimately leads to one Ly-$\alpha$ photon 
and one Balmer-$\alpha $ photon.

\subsubsection{Procedure ``B" : Modelling line emission from the gas}

A sophisticated scheme  of radiation transfer through the 
interstellar gas component has been used to explain the 
line emission from the gas. These include 
infrared nebular/ionic fine structure lines.
This procedure considers
several prominent elements in the gas phase of the cloud.
In addition to photoionization and recombination,
other physical processes like collisional excitation \& 
de-excitation, grain photoionization and gas-dust coupling
are also included.
This detailed modelling involves the
use of the photoionization code CLOUDY (Ferland, 1996),
which has been supplemented with a software scheme developed 
by Mookerjea \& Ghosh (1998), to make the model predictions more
realistic and easy to compare with observations.
This scheme improves the modelling 
by (i) emulating the exact structure of the H II
region; and (ii) including
 the absorption effects of the dust (present within the 
line emitting zones), on the emergent line intensities.
It predicts physical conditions of the gas, e.g. ionization, 
level populations, temperature structure, and the 
emerging emission line spectrum.
     The gas component of the cloud has been 
considered with typical H II
region abundance, as tabulated by Ferland (1996). 
Only the elements with abundance 
relative to hydrogen,  higher than $3.0\times 10^{-6}$ have been
used; these are -- H, He,
C, N, O, Ne, Mg, Si, S and Ar. The grains of the 
Astronomical Silicate
and Graphite types have been introduced at and beyond 
a radial distance from the exciting star such that they
do not heat up above their sublimation temperature.

The entire cloud is considered to be consisting of
two spherical shells, the inner one made of gas alone and
the outer one with gas and dust. The boundary between
the two shells,
R$_{\rm min}$, is taken from the corresponding 
best fit model using Procedure ``A''.
CLOUDY is run twice, the first time (RUN1) for the inner pure gas
shell with the central energy source. 
The resulting emergent continuum from RUN1 is used as input to
the second run (RUN2) for the outer shell.
The emerging line spectrum from RUN1 is transported 
to an outside observer,
through the second (outer) shell by considering the 
extinction due to the entire dust column present there.
For every spectral line considered, its emissivities 
from individual radial zones of RUN2
are transported through the corresponding remaining 
dust column densities within the outer shell.
The emerging line luminosities from RUN1 and RUN2 are
finally added to predict the total observable luminosity.

A total of 27 most prominent spectral lines (from various
ionization stages of the above mentioned 10 elements)
in the wavelength range 2.5 -- 200 \micron\ 
have been considered. From observational point of view, 
the reliable detectability of any spectral line will
depend on experimental detail like : 
the instrumental line profile (spectral resolution);
as well as the strength of the continuum
in the immediate spectral neighbourhood of the line. 

\subsection{IRAS 00338+6312}

This source has been modelled using  Procedure ``A" only, since 
no high resolution spectroscopic observation is available.
We present here the model, which was found to be the most consistent with 
all available observations.
The SED for IRAS 00338+6312 has been constructed from 
flux densities at the 4 IRAS bands (from HIRES maps) and the 2 TIFR 
bands. The flux densities at the sub mm
wavelengths of 800 and 1100 \micron\ have been taken from JCMT 
observations by McCutcheon et al. (1995).  
The near-IR data (Persi et al. 1988; Weintraub \& Kastner 1993)
have not been used in the SED, since now it is clear that
the RNO 1B/1C are not physically associated with IRAS 00338+6312.

The total color corrected luminosity of IRAS 00338+6312 as quoted
by Wouterloot \& Brand (1989)
is 3$\times 10^{3}$ \lsun\ for a distance of 1.61 kpc.
However more recent observations 
give the distance to the source to be 1.1 kpc, implying the total
luminosity to be $\sim$ 1056 \lsun\ (McCutcheon et al. 1995).
The latter number has been used by us for modelling purposes.
Snell, Dickman \& Huang (1990)
and Carpenter, Snell \& Schloerb (1990) have  determined
the cloud mass from their CO measurements, to be between 
2.2$\times 10^{3}$ \msun\
(M$_{LTE}$) and 2.7$\times 10^{3} $ \msun\ (M$_{virial}$).
Although the radio continuum observations of IRAS 00338+6312 by 
Carpenter, Snell \& Schloerb (1990) and McCutcheon et al. (1991)
at 6 cm failed detection, Anglada et al. (1994) successfully detected it
at 3.6 cm (0.49 mJy). This radio continuum measurement has been 
used to fine tune the gas to dust ratio in the model for this source.
The resulting best fit model is presented in Fig. 8.
Although the fit obtained is quite good at the infrared
wavelengths from 12 to 185 \micron, at the sub-mm wavelengths (800
\& 1100 \micron) the model predicts a large excess over the observed fluxes.
One possible explanation of this discrepancy is that, whereas the region
mapped by JCMT does not cover the entire cloud, our model
prediction displayed in Fig. 8 is the integrated emission from 
the entire cloud.  The radius of
the map shown by McCutcheon et al. (1995) is less than 0.3 pc, while 
other CO
observations give the radius of the source to be $\sim$ 1.5 pc. 
Our model predicts an outer radius of 1.1 pc. We have
estimated the total flux densities at 800 and 1100 \micron\ by
calculating the fraction of the total flux emitted 
from the area mapped by McCutcheon et al.,
from the model and corrected the observed fluxes accordingly. 
These extrapolated 
flux densities are quite close to the model predictions.
The best fit model parameters are summarised in Table 2.
The most surprising result from our modelling is that,
the dust is radially distributed uniformly ($n_{d}\sim r^{0}$).
The models with density distribution as
$r^{-1}$ and $r^{-2}$ yield extremely poor and unacceptable fits to the
observed SED. 
The embedded energy source is a single ZAMS B2.5 type star.
The inner dust radius (cavity size) is consistent with 
the dust sublimation at the relevant local radiation field.
The role of external boundary condition in the form of
assumed ISRF, is evident at wavelengths shorter than
$\approx$ 3 \micron\ in the SED. At these wavelengths,
the backscattered ISRF dominates over the thermal
dust emission.  
The total cloud mass corresponding to this model is
2.4 $\times 10^{3}$ \msun\ , which is in excellent
agreement with the mass determined from the CO data 
(2.2 -- 2.7 $\times 10^{3} $ \msun\ ; Carpenter, Snell \& Schloerb, 1990).
The expected angular sizes of IRAS 00338+6312 in the HIRES 
and TIFR maps have been estimated from the model, by convolving
the predicted source size (at a particular wavelength band) 
with the achieved angular resolution (post image processing) in that
band. These predictions are consistent with the 
sizes seen at 12, 25, 60, 100, 143 \& 185 \micron\ maps.

Hence, our model for IRAS 00338+6312 can be called very 
successful since it not only fits the observed infrared sub-mm SED, 
but also selfconsistently explains the radio continuum emission
and is consistent with the implications of the CO data.

\subsection{RAFGL 5111 (IRAS 03595+5110)}

Compared to the previous source (IRAS 00338+6312), 
modelling of RAFGL 5111 is
rather involved, since it has a complex geometry as implied by its
structure at widely different wavebands encompassing 
optical, infrared and radio wavelengths.
IRAS 03595+5110 is associated with an evolved H II region
with an optical nebula, S 206. This nebula appears to be open
or mass-limited in the eastern direction and photon limited
to the west (Deharveng, Israel \& Maucherat 1976). 
The geometry is similar
to the ``blister'' type of H II region, discussed by
Icke, Gatley \& Israel (1980), where a dense interstellar
cloud is heated by a young star situated just outside the
cloud (or near the boundary of the cloud). The position
of IRAS 03595+5110 is situated near the centroid of the
dense cloud, thereby representing the thermal mid \&
far infrared emission from the dust component. 
The distance to this source is 3.3 kpc (Snell, Dickman
\& Huang 1990).
The exciting
star (outside the cloud), has been identified to be 
BD +50${}^{0}$ 886. This star of spectral type O6, can explain
the energetics of IRAS 03595+5110, despite the geometrical
dilution arising due to its far location with respect to the cloud.
The corresponding ionized nebula has been studied well
at radio continuum as well as radio recombination lines.
The VLA map at 20 cm by Albert et al. (1986) shows structure
such that a 1-D cut in the east-west direction turns
out to be consistent with the ``blister'' type model 
with an exciting star with
luminosity $\approx$ 8.2$\times 10^{4}$ \lsun\ , and
$N_{Lyc} \approx$ 5.3$\times 10^{48}$ s$^{-1}$.
The detailed study of this source at 3.4 cm by Balser et al. (1995)
using interferometric (VLA; 8\arcsec\  beam) as well as 
single dish (100-m Effelsberg; 84\arcsec\  beam) measurements, has
led to determination of many parameters. The single dish 
map has been modelled for a homogeneous spherical shell geometry
as well as a spherical gaussian geometry. In either case,
an O6.5 ZAMS exciting star is implied and the electron 
density, $n_{e}$,  ranges between 180 -- 240 cm${}^{-3}$.
However, the compact component resolved in the VLA map,
when modelled, led to an O9.5 type of exciting star and
$n_{e} \approx$ 770 cm${}^{-3}$.
The CO line emission has been detected from this source
(Snell, Dickman \& Huang 1990), but no outflow activity
has been found. Searches for H$_{2}$O masers in the 
neighbourhood of this source have been unsuccessful. The
above two points are consistent with the fact that in
IRAS 03595+5110, one is dealing with a very evolved
H II region/star forming region.
The nebular ionic lines ([O III] at 52 \& 88 $\mu$m;
[N III] at 57 $\mu$m) in the far infrared
have been detected from this source by Rubin et al. (1988).

  The observed SED for IRAS 03595+5110 has been compiled from
the following measurements : IRAS HIRES maps (12, 25, 60 \& 100
\micron); TIFR maps (143 \& 185 \micron) and RAFGL data (11, 20 \&
27 \micron). The near-IR data of Pismis \& Mampaso (1991) has not
been used, since none of their peaks (IRS1/IRS2/IRS3) coincide
with the IRAS source, which is not surprising. In fact their
IRS1 is the exciting star BD +50${}^{0}$886.

 In spite of the complex geometry of IRAS 03595+5110, 
we have attempted to construct a self consistent
picture of this source using an ``equivalent''
spherically symmetric model. The artificial aspect in this
``equivalent'' model approach is that, an embedded energy source
is incorporated, although, in actual reality, the
cloud is heated/ionized/illuminated from outside. 
However, the problem can be made energetically ``equivalent''
by quantifying the luminosity intercepted in the actual
case and using that as the luminosity constraint on the
embedded source. 
Since, spectroscopic measurements of ionic lines are
available for IRAS 03595+5110, it has been modelled using
both Procedures ``A" and ``B". 

\subsubsection{Results from Procedure ``A''}

The derived parameters corresponding to the best fit
model from Procedure ``A" are presented in Table 2.
This model corresponds to a uniform
the dust density distribution ($n(r) \sim r^{0}$). 
The embedded energy source is
a single ZAMS O6.5 star with 52\% of its total luminosity 
deposited in the cloud (7.7$\times 10^{4} $ \lsun\ ).
This special situation is demanded by the simultaneous
fitting of the SED as well as the radio continuum emission.
The model fit to the SED is shown in Fig. 8.
The total radio continuum emission predicted by this model is 1.5
Jy at 8.75 GHz, which falls in between the values obtained from 
interferometric and single dish observations (Balser et al. 1995).
The model predicts the electron density inside the H II region,
$n_{e}$, to be 530 cm${}^{-3}$, which is again within the range of
values obtained by Balser et al. (1995) from the VLA (770
cm${}^{-3}$) and single dish data (240 cm${}^{-3}$).
The cloud mass from the CO observations (Snell, Dickman \&
Huang 1990) of the central 2\arcmin\ (dia) turns out to be
$\approx 69 $ \msun\ , which is similar to the mass of the
central 2\arcmin\ of the model cloud, $46 $ \msun\ .
 Considering the complex geometry of IRAS 03595+5110, the above 
model can be called quite satisfactory.

\subsubsection{Results from Procedure ``B''}

 The modelling for IRAS 03595+5110
using the Procedure ``B'', has been carried out using
the physical sizes and other parameters from the
best fit model under scheme ``A''.
  In all 18 nebular/ionic lines satisfied our detectability
criterion (which is the following : the power in the line 
is at least 1\% of the power in neighbouring continuum
for a resolution element determined by $R = \lambda / {\Delta
\lambda}$ = 1000). The wavelengths and luminosities of these lines 
are presented in Table 3.
The complete emerging spectrum predicted by this model,
including all lines from the 10 elements considered and 
the continuum, is shown in Fig. 9.

  A comparison of observed line ratios, \namely ,
[O III (52 \micron)] / [O III (88 \micron)] and  
[N III (57 \micron)] / [O III (52 \micron)]
(Rubin et al. 1988), with our model predictions are shown in Table 4.
Whereas the former ratio is overestimated in the model, the latter is
quite close to the measurements.
The model predictions can be much better tested if this source
is studied in many other infrared lines, by present/future space 
missions (ISO, SOFIA, SIRTF).

  It is interesting to note that the 
mid to far infrared continuum levels predicted from
the best fit models for RAFGL 5111 by Procedures ``A" and ``B"
match within 20 -- 30\%. This is remarkable, considering
the diverse approaches of the two procedures.

\subsection{IRAS 04004+5114}

   This faint source in the neighbourhood of RAFGL 5111,
which was also mapped at the two TIFR bands,
has been considered for radiative transfer modelling
using Procedure ``A''.
Its observed SED has been constructed from the IRAS HIRES
data (at 12, 25, 60 \& 100 \micron) and the TIFR data
(at 143 \& 185 \micron).
The distance to IRAS 04004+5114 has been taken to be 
2.6 kpc (Wouterloot, Brand \& Fiegle 1993).

  The spherically symmetric dust cloud model fitting the
observed SED best, corresponds to : a constant density 
cloud (i.e. $n(r) \sim r^{0}$); an embedded star of type
A3 with total luminosity 1.7$\times 10^{3} $ \lsun. 
Other derived parameters are listed in Table 2.
The model fit to the SED for IRAS 04004+5114 is shown
in Fig. 8.  It can be seen that the model fits the 
available observations very well. The predicted radio emission 
for this source is 0.1 mJy at 5 GHz.

\section{Summary}

  Two Galactic star forming regions associated with the
IRAS sources 00338+6312 and 03595+5110 (RAFGL 5111)
have been studied
in detail. The source IRAS 00338+6312, is very controversial
since earlier it was thought to be excited by 
FU Orionis type activity, though only recently evidences
showed that it is a very young and deeply embedded 
object in a much earlier
evolutionary phase of star formation. 
The second source is associated with an evolved H II region
in a molecular cloud (S 206), excited by a star located outside
the cloud itself (an example of ``blister'' type geometry). 
Although both these sources have red
nebular objects in their IRAS error ellipses, they
represent probably two extremes in stage of evolution
of star forming regions.

Both these sources have been mapped (angular resolution 
$\sim$ 1\farcm 5) simultaneously in two
far infrared bands ($\lambda_{\rm eff}$ = 143 \& 185 \micron),
using the TIFR 100 cm balloon borne telescope. 
The HIRES processed IRAS survey data for the same region 
at 12, 25, 60 \& 100 \micron have also been used for comparison.

Whereas IRAS 00338+6312 is only slightly extended (and hence
resolved only in the TIFR maps), IRAS
03595+5110 is well resolved in both the TIFR bands, as also
in some IRAS bands. The faint source in the neighbourhood of
RAFGL 5111, \namely, IRAS 04004+5114, has also been resolved
in the TIFR bands.
The dust colour temperature, $T(143/185)$, and optical depth,
$\tau_{\rm 150}$, maps have been generated
for RAFGL 5111. Similarly the $T(60/100)$, and $\tau_{100}$
maps have been generated from IRAS data for comparison. 
These maps show interesting structures implying dust density and/or 
temperature gradient along a 
line joining the external exciting star and the outer boundary
of the cloud.

 Two independent radiative transfer modelling procedures have been 
implemented in spherical geometry, for a dust-gas cloud  with 
an embedded energy source at the centre. 
The observed SED, radio continuum data, infrared fine structure 
line strengths etc. have been used to optimise these models.
The resulting models for IRAS 00338+6312, 03595+5110 and
04004+5114 have been presented. 
The most important conclusion is that for all the above
three sources, the best fit models correspond to the 
uniform radial density distribution law ($n(r) \sim r^0$).
Intuitively one would expect a centrally peaked radial
density distribution in star forming regions.
Identical conclusion has been found in the studies of 
several similar
Galactic star forming regions (Faison et al. 1998, Campbell et al. 1995,
Colome et al. 1995, Butner et al. 1994; Mookerjea \&
Ghosh 1998). The model predictions of nebular line intensities
for RAFGL 5111, are in reasonable agreement with observations.
  
\centerline{\bf Acknowledgements}

 We would like to thank IPAC (Caltech) for providing the HIRES
processed IRAS data.
Gary Ferland is thanked for his help on many occasions
regarding the code CLOUDY.
  It is a pleasure to thank  S L D'Costa, 
M V Naik,
D M Patkar,
M B Naik,
S A Chalke,
S V Gollapudi,
G S Meshram and
C B Bakalkar of the Infrared Astronomy Group for their
technical support to the Far Infrared Astronomy Programme.
Thanks are due to M N Joshi, S Sreenivasan, J V Subbarao and
other colleagues of TIFR Balloon Facility at Hyderabad for
smoothly conducting the balloon flight and related operations.
We also thank  all members of the Control Instrumentation (CIBA) Group.

We thank the anonymous referee and Steven Willner 
(Scientific Editor) for their suggestions which improved the 
presentation of the paper.





\newpage

\figcaption[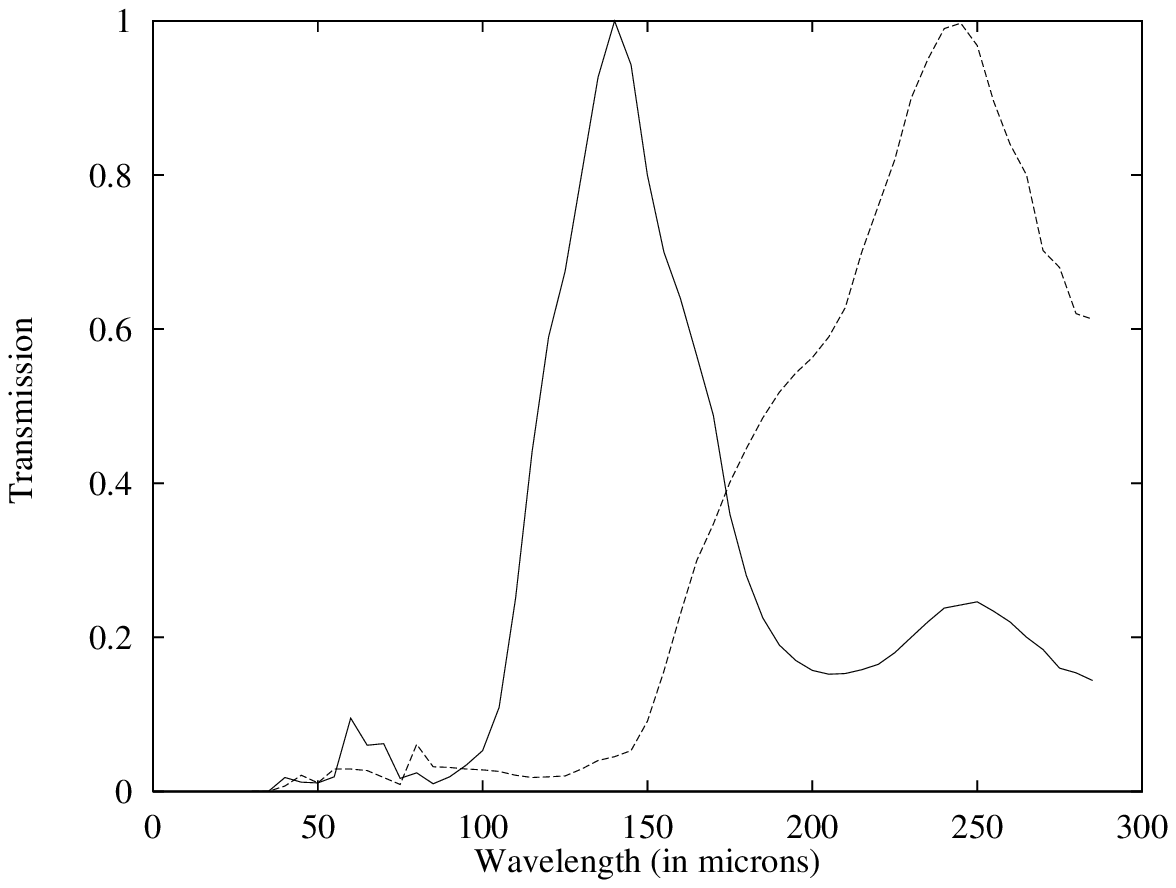]
{Normalized spectral responses of the two far infrared bands of the
TIFR photometer system. The solid line corresponds to the band with
$\lambda_{eff}$ = 143 \micron\ and the dashed line  corresponds to the band
with $\lambda_{eff}$ = 185 \micron .
\label{fig:f1}
}

\figcaption[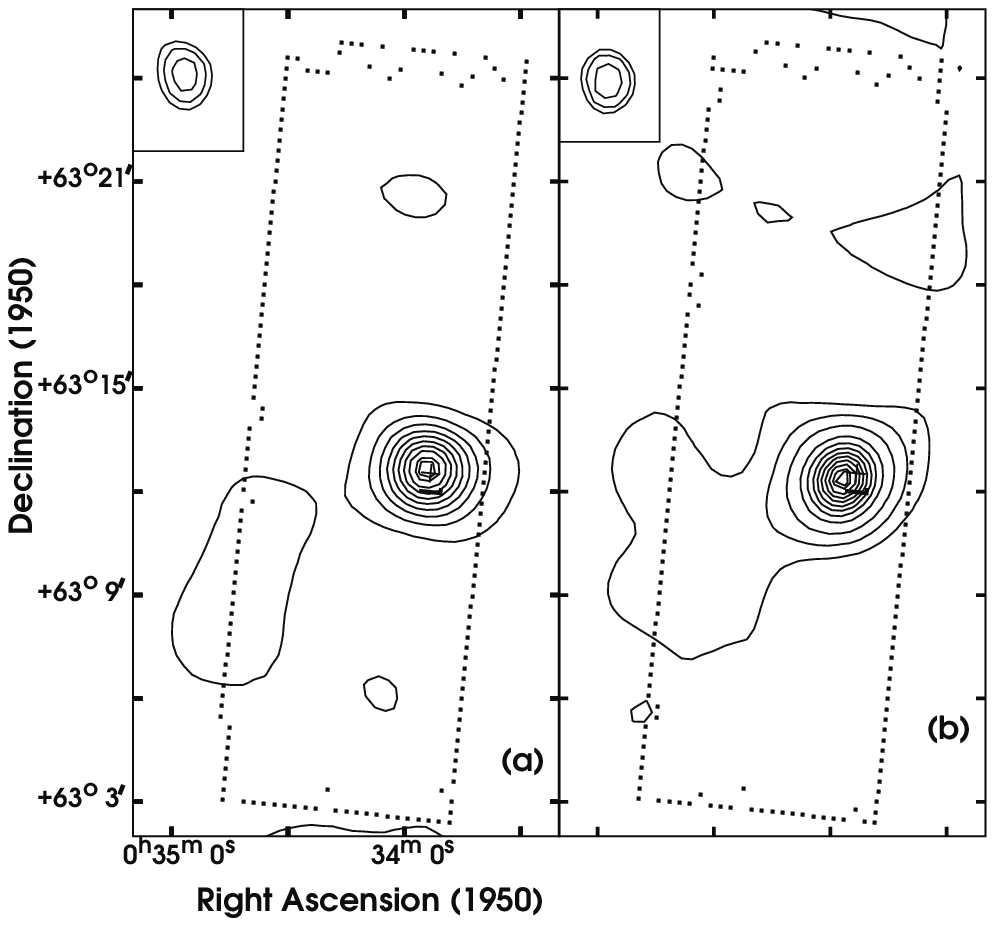]
{TIFR Intensity maps for IRAS 00338+6312 : (a) at 143 \micron\ the
Peak = 368.0 Jy/ sq. arcmin ; Contour levels = 10, 20, 30, 40, 50, 60, 70, 
80, 90, 95\% of the peak; (b) at 185 \micron\ the Peak = 809.0 Jy/ sq. arcmin ;
Contour levels = 2.5, 5.0, 10, 20, 30, 40, 50, 60, 70, 80, 90, 95\% of the peak.
The insets show deconvolved images of Saturn in the respective bands. The
contours are 50, 70 and 90 \% of respective peaks, aligned to the instrumental
axes for meaningful comparison.  ``+'' shows the IRAS PSC position of 
the source. 
\label{fig:f2}
}

\figcaption[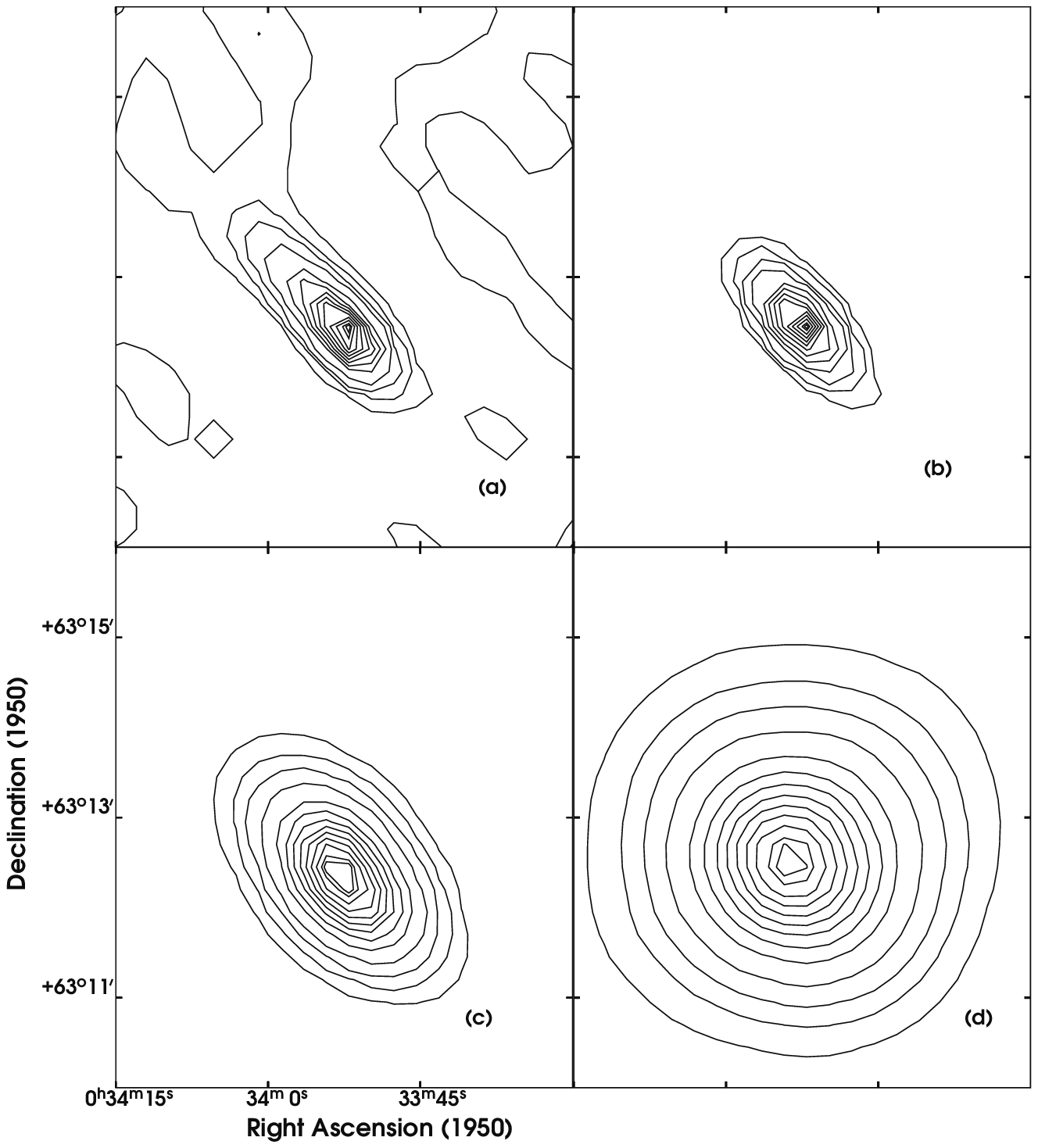]
{HIRES processed IRAS maps of IRAS 00338+6312: (a) at 12 \micron , (b) 25 \micron , (c) 60 \micron\ and (d) 100 \micron\ with  peaks = 8.2, 82.8, 312 and 284 Jy/sq. arcmin respectively.   The contour levels are 1, 2.5, 5, 10, 20, 30, 40, 50, 60, 70, 80, 90 \& 95 \% of the peaks in corresponding bands.
\label{fig:f3}
}

\figcaption[fig4]
{(a) \& (b) show TIFR maps for IRAS 03595+5110  at 143 \micron, 
(Peak = 142.0 Jy/sq. arcmin) and  at 185 $ \mu$m (Peak = 99.2 
Jy/sq. arcmin) respectively.  The contour 
levels are the same as in Fig. 2(a). The insets are identical to those 
in Fig. 2.  ``+'' shows the IRAS PSC position of the source. (c), (d), (e) 
\&  (f) show 
HIRES processed IRAS maps of the same source at 12 \micron , 25 \micron, 
60 \micron\ and 100 \micron\ with  peaks = 9.66, 93.4, 237 and 130 Jy/sq. arcmin
respectively. The contour levels are the same as in Fig. 3.
\label{fig:f4}
}

\figcaption[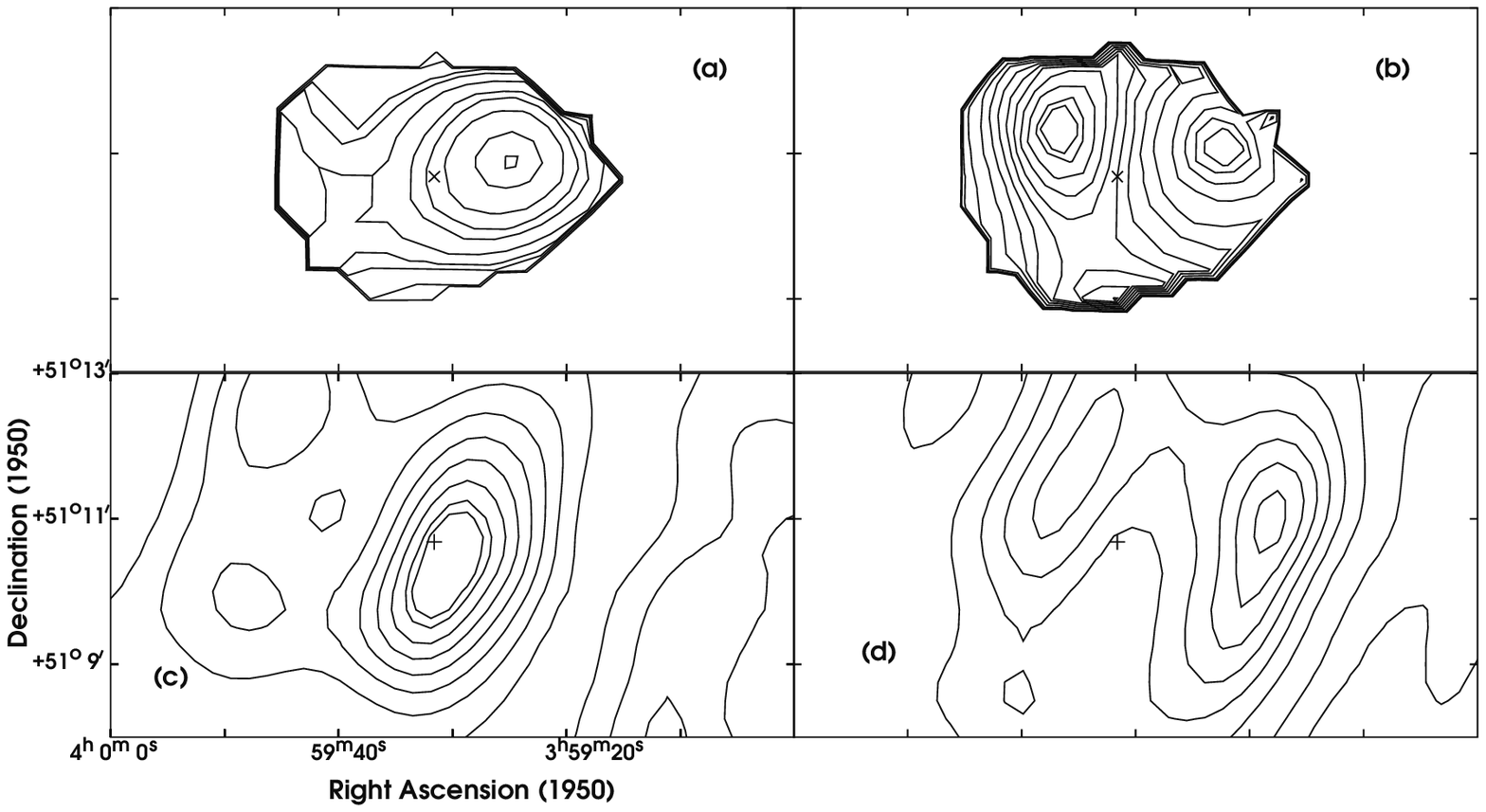]
{(a) \& (c) show the $T(143/185)$ and $T(60/100)$ maps respectively of the 
source IRAS 03595+5110. The contours in (a) are at 20, 22, 24, 25, 27, 30, 
32, 35, 40 \& 43 K. The contours in (c) are at 24, 25, 26, 28, 33, 37, 42, 51,
55, 60, 64, 67 K. (b) \& (d) show the  $\tau_{\rm 150}$ and \dep\
maps repectively. The peak in (b) is 2.7$\times$10$^{-3}$ and the contours 
shown are
the same fractions of the peak as in Fig. 2(b). The peak in (d)
is 1.25$\times$10$^{-3}$ and the contours are same fractions of the 
peak as in Fig. 2(a). ``+'' shows the IRAS PSC position of 
the source. All the maps have been generated assuming 
$\epsilon_{\lambda}\sim\lambda^{-1}$ emissivity law.
\label{fig:f5}
}

\figcaption[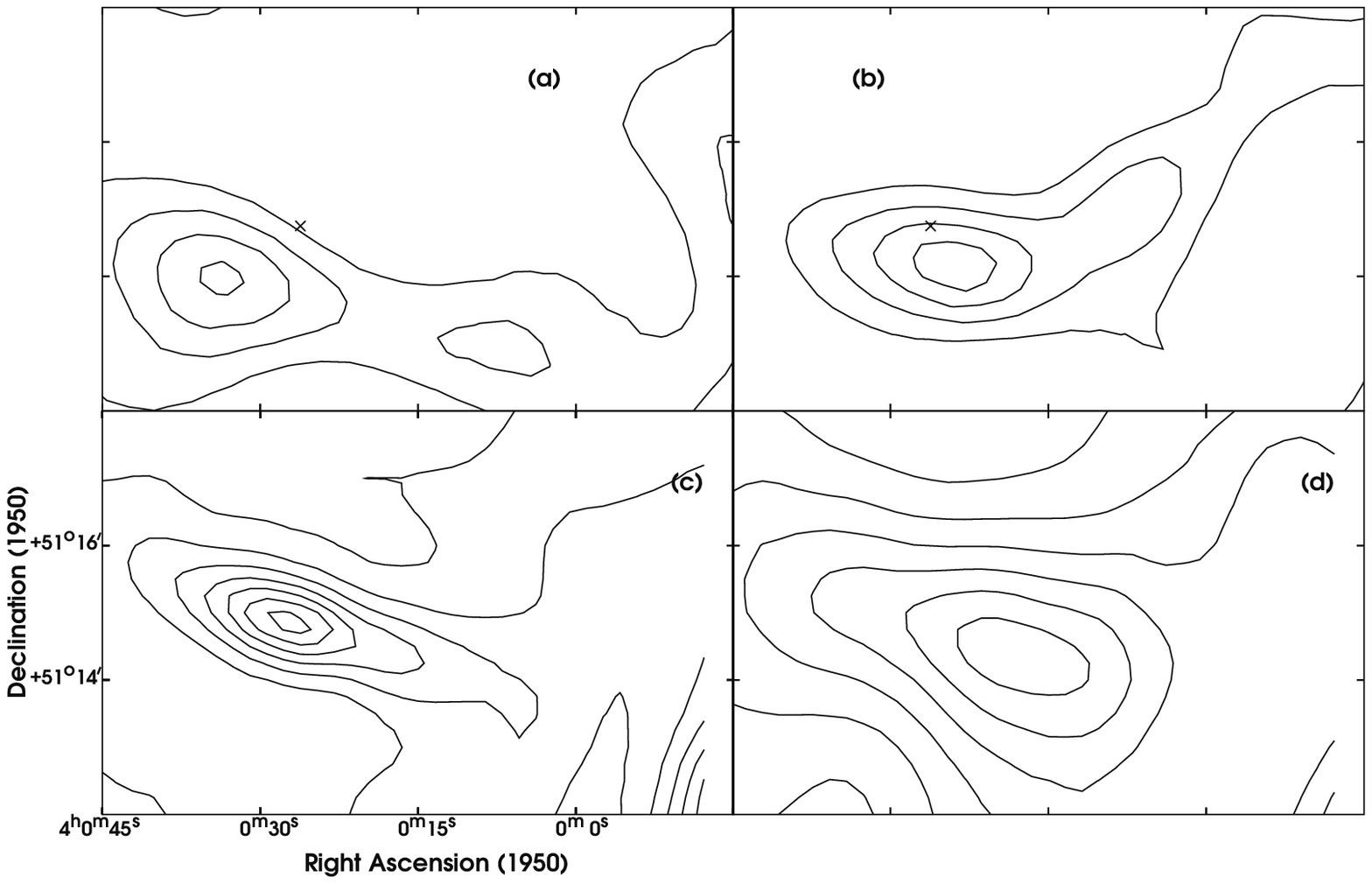]
{(a) \& (b) show specially processed (see text) TIFR maps for IRAS 04004+5114  
at 143 \micron\ (Peak = 34.3 Jy/sq. arcmin) and 185 \micron\ (Peak = 31.7 
Jy/sq. arcmin) respectively. Contour levels = 40, 60,80, 100\% of the 
respective peaks. The ``+'' shows the IRAS PSC position of the source.
(c) \& (d) show HIRES maps of the same source, at 60\micron\ (Peak = 22.7 Jy/sq.
arcmin) and 100 \micron\ (Peak = 23.8 Jy/sq. arcmin) respectively. The
contour levels are the same as in Fig.3
\label{fig:f6}
}

\figcaption[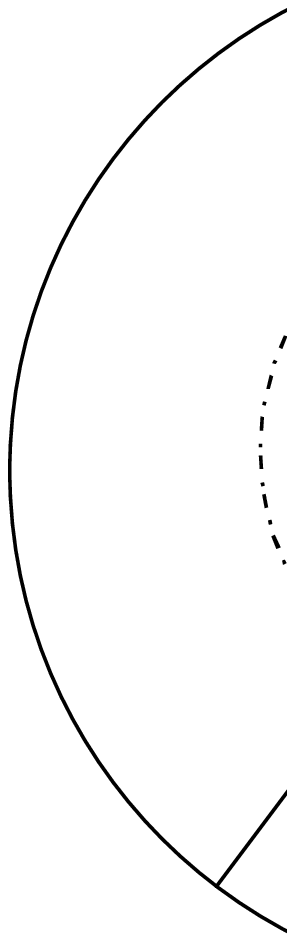]
{Schematic diagram of the interstellar cloud with an embedded star. 
\label{fig:f7}
}

\figcaption[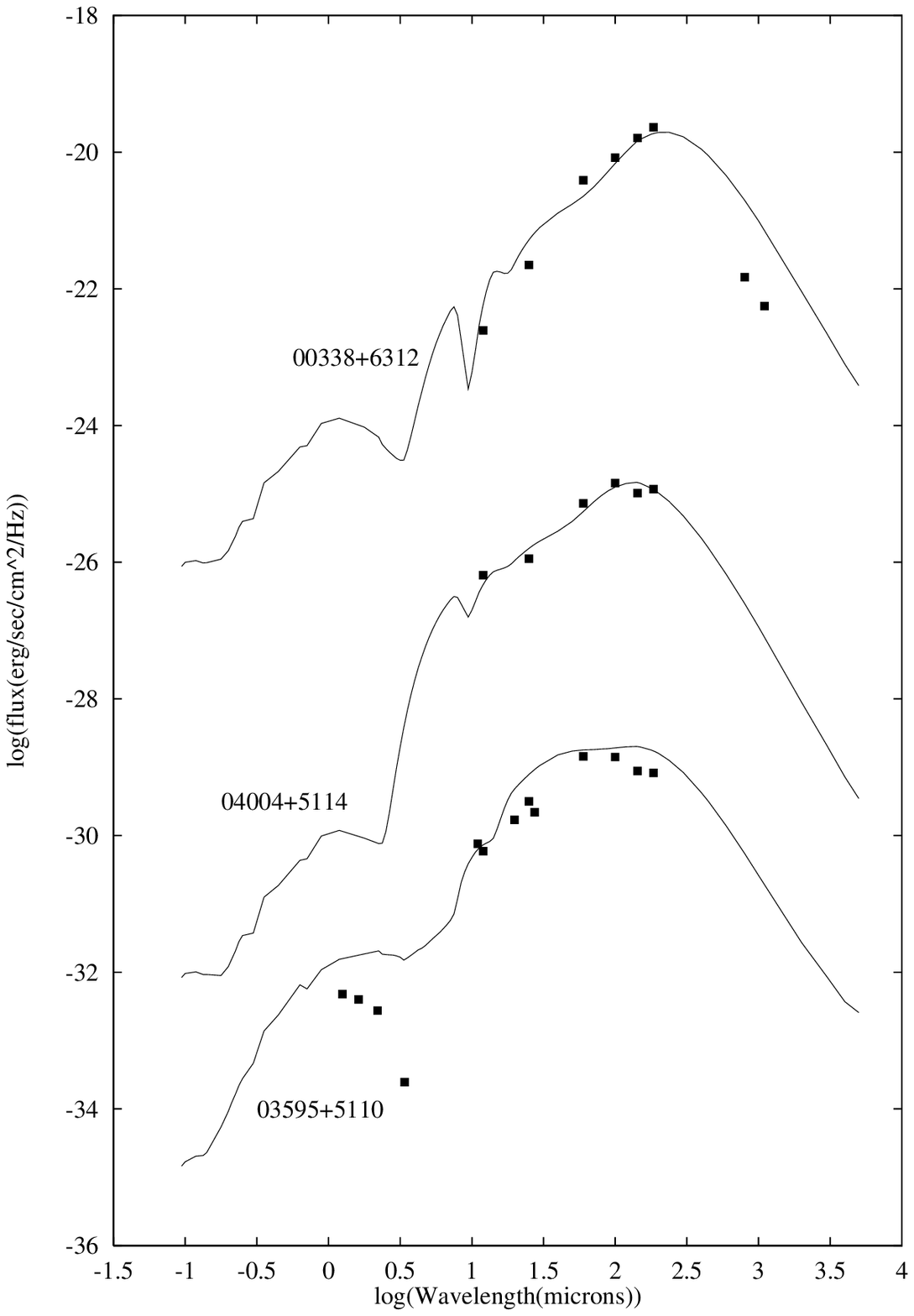]
{Spectral energy distributions for  IRAS 00338+6312, IRAS 04004+5114 and
IRAS 03595+5110. For all the sources the continuous line represents the 
best fit radiation transfer model (procedure ``A'') and the points 
represent observations. For better visibility the SEDs of IRAS 04004+5114
and IRAS 03595+5110 have been shifted by -4  and -9 dex respectively.
\label{fig:f8}
}

\figcaption[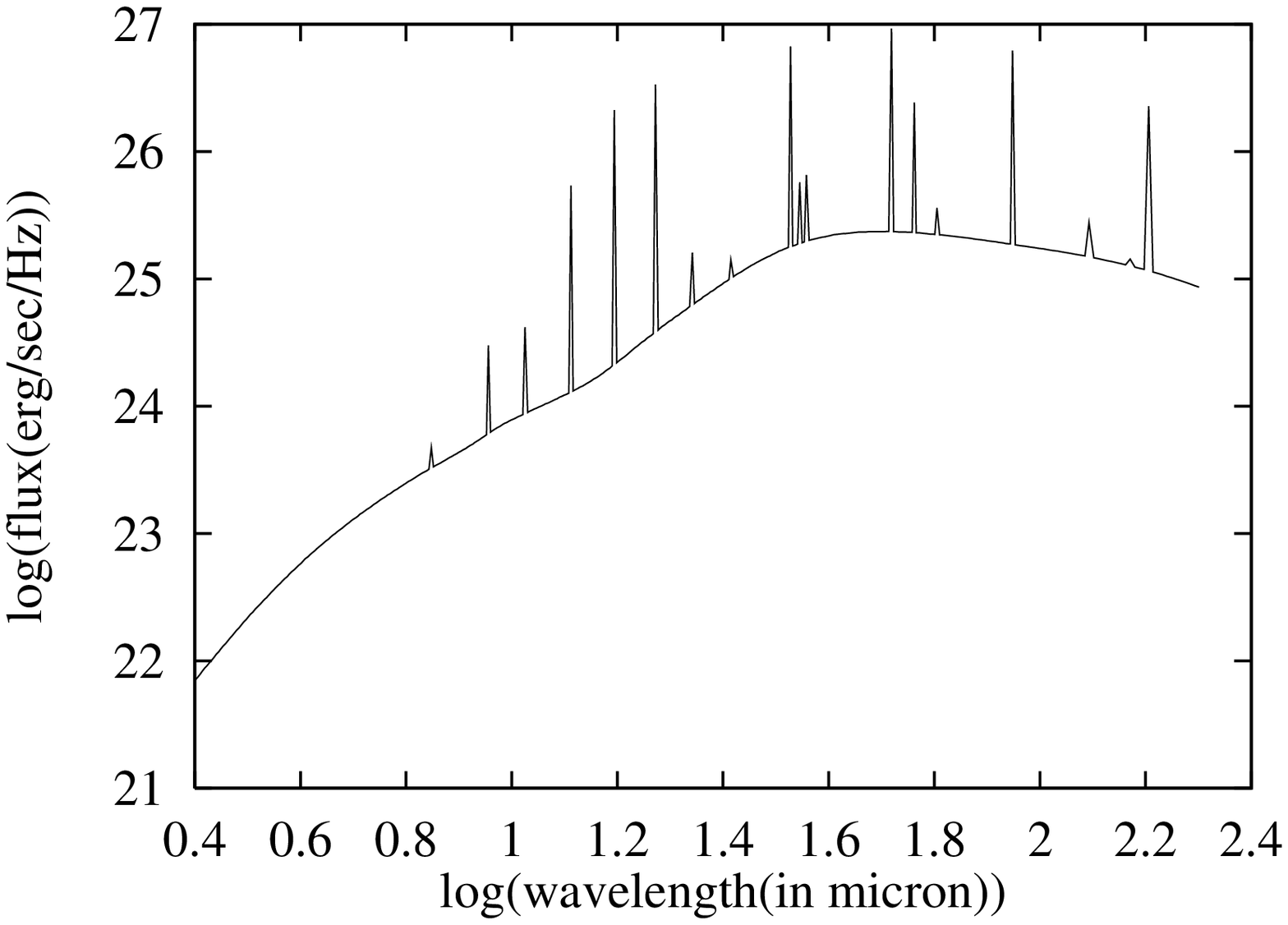]
{Emergent spectrum from model calculations of IRAS 03595+5110, using 
procedure ``B''. 
\label{fig:f9}
}




\newpage

\begin{deluxetable}{ccccccccc}
\tablecaption{Positions (143 \micron\ or 185 \micron ) and flux densities of the sources from TIFR and IRAS HIRES maps.}
\tablehead{
\multicolumn{1}{l}{Source} & 
\multicolumn{2}{c}{Position} & 
\multicolumn{1}{c}{F$_{\rm {12}  \mu \mathrm {m} }$ }& 
\multicolumn{1}{c}{F$_{\rm {25}  \mu \mathrm {m} }$ }& 
\multicolumn{1}{c}{F$_{\rm {60}  \mu \mathrm {m} }$ }& 
\multicolumn{1}{c}{F$_{\rm {100} \mu \mathrm {m} }$} & 
\multicolumn{1}{c}{F$_{\rm {143} \mu \mathrm {m} }$} & 
\multicolumn{1}{c}{F$_{\rm {185} \mu \mathrm {m} }$}\\
& \colhead {R.A. (1950)} &
\colhead {Dec. (1950)}&
\colhead {Jy}&
\colhead {Jy}&
\colhead {Jy}&
\colhead {Jy}&
\colhead {Jy}&
\colhead {Jy}
}
\startdata
00338+6312 & 00$^{\rm h}$33$^{\rm m}$53.$^{\rm s}$3 & +63$^{\rm \circ}$12\arcmin32\arcsec
 & \phn 2.5 & \phn 22.4 & \phn 389 & \phn 829 & 1615 & 2317 \nl
(5 \arcmin\ dia)&&&&&&&& \nl
03595+5110 & 03$^{\rm h}$59$^{\rm m}$31.$^{\rm s}$6  &+51$^{\rm \circ}$10\arcmin 41\arcsec
 &  58.6 & 314.7 & 1462 & 1405 & \phn 883 & \phn 829 \nl
(5 \arcmin\ dia)&&&&&&&& \nl
04004+5114 &04$^{\rm h}$00$^{\rm m}$26.$^{\rm s}$1  &+51$^{\rm \circ}$14\arcmin 45\arcsec
& \phn 6.4 & \phn 11.3 & \phn \phn 72 & \phn 146 & \phn 100 & \phn 118 \nl
(3 \arcmin\ dia)&&&&&&&& \nl
\enddata
\end{deluxetable}

\begin{deluxetable}{ccccccccc}
\tablecaption{Best fit parameters for the three sources modelled, using Procedure ``A''.}
\tablewidth{0pt}
\tablehead{
\multicolumn{1}{c}{Source} & 
\multicolumn{1}{c}{R$_{\rm max}$} & 
\multicolumn{1}{c}{R$_{\rm min}$ }& 
\multicolumn{1}{c}{ \dep\ }&
\multicolumn{1}{c}{Luminosity} & 
\multicolumn{1}{c}{Silicate : } & 
\multicolumn{1}{c}{N$_{e}$ }&
\multicolumn{1}{c}{M$_{\rm Dust}$} & 
\multicolumn{1}{c}{Gas:Dust }\\
& \colhead{(pc)} & 
\colhead{(pc)} & & 
\colhead{(10$^{3}$ \lsun\ )}& 
\colhead{ Graphite} & 
\colhead{(cm${}^{-3}$)} & 
\colhead{( \msun\ )}  
}
\startdata
00338+6312 &  1.1\phn & 0.0001 & 0.15\phn \phn & \phn 1.8 & 31.9:68.1 & 1.8$\times 10^{4}$ & 47.2\phn & 50:1 \\
& & & & & & & & \\
03595+5110 &  5.0\phn & 0.002\phn & 0.0095 & 77.0 & 98.3:1.7 & 530 & 65.22 & 100:1\\
& & & & & & & & \\
04004+5114 &  0.25 & 0.0003 & 0.08 \phn \phn & \phn 1.7  &  21.4:78.6 & 8.4$\times 10^{4}$ & 129.2\phn & 100:1 \\
\enddata
\end{deluxetable}

\begin{deluxetable}{lcc}
\tablecaption{Emergent line luminosities predicted by the model for IRAS 03595+5110, using Procedure ``B''.}
\tablehead{
\multicolumn{1}{l}{Element \&} & 
\multicolumn{1}{l}{Wavelength }& 
\multicolumn{1}{l}{Luminosity } \\
\colhead{  Ionization stage} &
\colhead{( $\mu$m)}&  
\colhead{($L_{\odot}$)}  
}
\startdata
C II & 157.74 &\phn 15.03 \\
O I &145.6\phn &\phn \phn 0.12 \\
N I & 121.8\phn &\phn \phn 1.15\\
O III &\phn 88.42 &\phn 75.25\\
O I &\phn 63.23 &\phn \phn 2.02\\
N III &\phn 57.26 &\phn 35.50\\
O III &\phn 51.85 & 161.50 \\
Ne III &\phn 36.04 &\phn \phn4.75 \\
Si II &\phn 34.84 &\phn \phn4.10\\
S III &\phn 33.50 &\phn 73.00\\
O IV &\phn 25.91 &\phn \phn0.57         \\
Ar III &\phn 21.84 &\phn \phn1.69\\
S III &\phn 18.69 &\phn 66.75\\
Ne III &\phn 15.57 &\phn 50.75\\
Ne II &\phn 12.82 &\phn 15.43\\
S IV &\phn 10.52 &\phn 23.48\\
Ar III &\phn \phn 8.99 &\phn 19.99\\
Ar II &\phn \phn 7.00 &\phn \phn 1.57\\
\enddata
\end{deluxetable}

\begin{deluxetable}{ccc}
\tablecaption{Comparison of model predictions of line ratios with observations.}
\tablehead{
\multicolumn{1}{c}{Lines}  & 
\multicolumn{1}{c}{Observations\tablenotemark{a}} & 
\multicolumn{1}{c}{Model (Procedure ``B'')} 
}
\startdata
[O III (52\micron)] / [O III (88\micron)] &  1.38 $\pm$ 0.07\phn & 2.15\\
& & \\
$[$ NIII (57\micron)$]$/ $[$ OIII (52\micron)$]$ &  0.18 $\pm$ 0.013 & 0.22\\
& & \\
\enddata
\tablenotetext{a}{Rubin et al. (1988)}
\end{deluxetable}

\end{document}